# A Case for Cooperative and Incentive-Based Coupling of Distributed Clusters*

Rajiv Ranjan, Aaron Harwood and Rajkumar Buyya
Grids Lab and P2P Networks Research Group
Department of Computer Science and Software Engineering
University of Melbourne, Victoria, Australia
{rranjan,aharwood,raj}@cs.mu.oz.au

October 31, 2018

## Abstract

Research interest in Grid computing has grown significantly over the past five years. Management of distributed resources is one of the key issues in Grid computing. Central to management of resources is the effectiveness of resource allocation as it determines the overall utility of the system. The current approaches to superscheduling in a grid environment are non-coordinated since application level schedulers or brokers make scheduling decisions independently of the others in the system. Clearly, this can exacerbate the load sharing and utilization problems of distributed resources due to suboptimal schedules that are likely to occur. To overcome these limitations, we propose a mechanism for coordinated sharing of distributed clusters based on computational economy. The resulting environment, called *Grid-Federation*, allows the transparent use of resources from the federation when local resources are insufficient to meet its users' requirements. The use of computational economy methodology in coordinating resource allocation not only facilitates the QoS based scheduling, but also enhances utility delivered by resources. We show by simulation, while some users that are local to popular resources can experience higher cost and/or longer delays, the overall users' QoS demands across the federation are better met. Also, the federation's average case message passing complexity is seen to be scalable, though some jobs in the system may lead to large numbers of messages before being scheduled. Our simulations show that the user population profile comprising 70% seeking optimize for cost and 30% seeking optimize for time, is a good population mix for the proposed system, as in this case every resource owner in the federation earns significant incentive. Further the total message count in this case is much lower when compared to other population profiles having greater percentage of the users seeking optimize for time.

## 1 Introduction

Clusters of computers have emerged as mainstream parallel and distributed platforms for high-performance, high-throughput and high-availability computing. Grid[23] computing extends the cluster computing idea to wide-area networks. A grid consists of cluster resources that are usually distributed over multiple administrative domains, managed and owned by different organizations having different resource management policies. With the large scale growth of networks and their connectivity, it is possible to couple these cluster resources as a part of one large grid system. Such large scale resource coupling and application management is a complex undertaking, as it introduces a number of challenges in the domain of security, resource/policy heterogeneity, resource discovery, fault tolerance, dynamic resource availability and underlying network conditions.

The resources on a Grid (e.g. clusters, supercomputers) are managed by local resource management systems (LRMSes) such as Condor[32] and PBS[11]. These resources can also be loosely coupled to form campus Grids using multi-clustering systems such as SGE[26] and LSF[2] that allow sharing of clusters owned by the same organization. In other words, these systems do not allow their combination similar to autonomous systems, to create an environment for *cooperative federation* of clusters, which we refer as Grid-Federation.

Scheduling jobs across resources that belong to distinct administrative domains is referred to as *superscheduling*. Majority of existing approaches to superscheduling[37] in a grid environment are non-coordinated. Superschedulers such as Nimrod-G[3], Tycoon[31], Condor-G[25], and GridBus-Broker[42] perform scheduling related activities independent of the other superschedulers in the system. They directly submit their applications to the underlying resources *without* taking into account the current load, priorities, utilization scenarios of other application

---

*extended version of the conference paper published at IEEE Cluster'05, Boston, MA

level schedulers. Clearly, this can lead to over-utilization or a bottleneck on some valuable resources while leaving others largely underutilized. Furthermore, these superschedulers do not have a co-ordination mechanism and this exacerbates the load sharing and utilization problems of distributed resources because sub-optimal schedules are likely to occur.

Furthermore, end-users or their application-level superschedulers submit jobs to the LRMS without having knowledge about response time or service utility. Sometimes these jobs are queued for relatively excessive times before being actually processed, leading to degraded QoS. To mitigate such long processing delay and enhance the value of computation, a scheduling strategy can use priorities from competing user jobs that indicate varying levels of importance. This is a widely studied scheduling technique (e.g. using priority queues)[5]. To be effective, the schedulers require knowledge of how users value their computations in terms of QoS requirements, which usually varies from job to job. LRMS schedulers can provide a feedback signal that prevents the user from submitting unbounded amounts of work.

Currently, system-centric approaches such as Legion[17, 44], NASA-Superscheduler[38], Condor, Condor-Flock[12], Apples[10], Punch[30], PBS and SGE provide limited support for QoS driven resource sharing. These system-centric schedulers, allocate resources based on parameters that enhance system utilization or throughput. The scheduler either focuses on minimizing the response time (sum of queue time and actual execution time) or maximizing overall resource utilization of the system and these are not specifically applied on a per-user basis (user oblivious). System centric schedulers treat all resources with the same scale, as if they are worth the same and the results of different applications have the same value; while in reality the resource provider may value his resources differently and has a different objective function. Similarly, a resource consumer may value various resources differently and may want to negotiate a particular price for using a resource. Hence, resource consumers are unable to express their valuation of resources and QoS parameters. Furthermore, the system-centric schedulers do not provide any mechanism for resource owners to define what is shared, who is given the access and the conditions under which sharing occurs[24].

## 1.1 Grid-Federation

To overcome these shortcomings of non-coordinated, system-centric scheduling systems, we propose a new distributed resource management model, called Grid-Federation. Our Grid-Federation system is defined as a large scale resource sharing system that consists of a coordinated federation (the term is also used in the Legion system and should not be confused with our definition), of distributed clusters based on policies defined by their owners (shown in Fig.1). Fig.1 shows an abstract model of our Grid-Federation over a shared federation directory. To enable policy based transparent resource sharing between these clusters, we define and model a new RMS system, which we call Grid Federation Agent (GFA). Currently, we assume that the directory information is shared using some efficient protocol (e.g. a peer-to-peer protocol[33, 27]). In this case the P2P system provides a decentralized database with efficient updates and range query capabilities. Individual GFAs access the directory information using the interface shown in Fig.1, i.e. subscribe, quote, unsubscribe, query. In this paper, we are not concerned with the specifics of the interface (which can be found in[35]) although we do consider the implications of the required message-passing, i.e. the messages sent between GFAs to undertake the scheduling work.

Our approach considers the emerging computational economy metaphor[3, 41, 43] for Grid-Federation. In this case resource owners: can clearly define what is shared in the Grid-Federation while maintaining a complete autonomy; can dictate who is given access; and receive incentives for leasing their resources to federation users. We adopt the market based economic model from[3] for resource allocation in our proposed framework. Some of the commonly used economic models[13] in resource allocation includes the commodity market model, the posted price model, the bargaining model, the tendering/contract-net model, the auction model, the bid-based proportional resource sharing model, the community/coalition model and the monopoly model. We focus on the commodity market model[45]. In this model every resource has a price, which is based on the demand, supply and value in the Grid-Federation. Our Economy model driven resource allocation methodology focuses on: (i) optimizing resource provider's objective functions, (ii) increasing end-user's perceived QoS value based on QoS level indicators[35] and QoS constraints.

The key contribution of the paper includes our proposed new distributed resource management model, called Grid-Federation, which provides: (i) a market-based grid superscheduling technique; (ii) decentralization via a shared federation directory that gives site autonomy and scalability; (iii) ability to provide admission control facility at each site in the federation; (iv) incentives for resources owners to share their resources as part of the federation; and (v) access to a larger pool of resources for all users. In this paper, we demonstrate, by simulation, the feasibility and effectiveness of our proposed Grid-Federation.

The rest of the paper is organized as follows. In Section 2 we summarize our Grid-Federation and Section 3 deals with various experiments that we conducted to demonstrate the utility of our work. Section 4 explores various related projects. We end the paper with some concluding remarks and future work in section5.

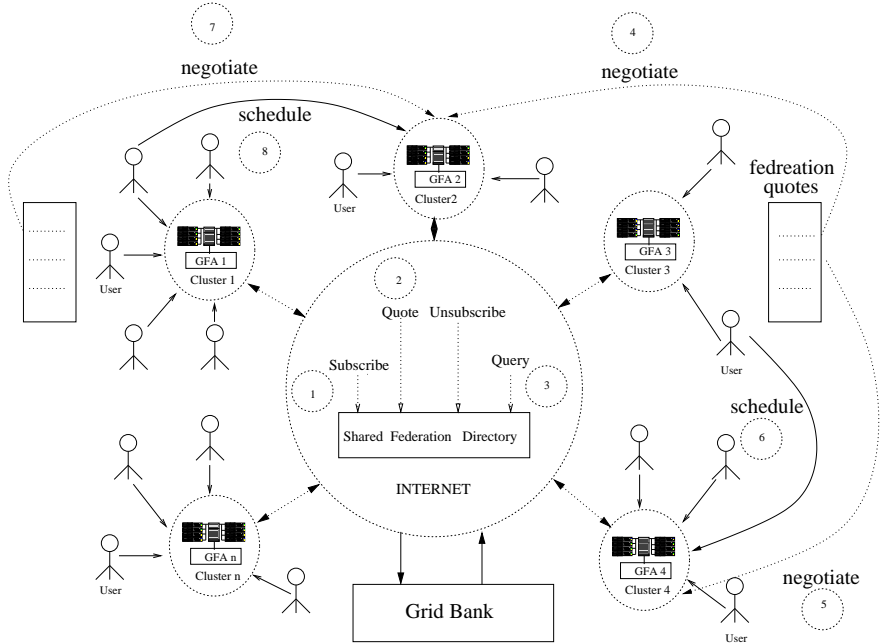

Figure 1: Grid-Federation

## 2 Grid-Federation models

This section provides comprehensive details about our proposed Grid-Federation, including models used for budget and deadline calculations in the simulations of the next section.

#### 2.0.1 Terms and Definitions

A *machine* is a single or multiprocessor system with memory, I/O facilities and an operating system. In this paper we define a *cluster* as a collection of homogeneous machines that are interconnected by a high-speed network like megabit or gigabyte Ethernet[28]. These machines work as integrated collection of resources. They have a single system image spanning over all the machines. A *resource management system* is a entity which manages a set of resources. The RMS can optimize any of the system-centric or user-centric requests on the underlying resources.

#### 2.0.2 LRMS(Cluster RMS)

In our proposed framework, we assume that every cluster has a generalized RMS, such as a SGE or PBS, that manages cluster wide resource allocation and application scheduling. Most of the available RMS packages have a centralized organization similar to the master-worker pool model. In the centralized organization, there is only one scheduling controller (master node) which coordinates system-wide decisions.

#### 2.0.3 Grid Federation Agent

We define our Grid-Federation (shown in Fig.1) as a mechanism that enables logical coupling of cluster resources. The Grid-Federation supports policy based[18] transparent sharing of resources and QoS[29] based job scheduling. We also propose a new computational economy metaphor for cooperative federation of clusters. Computational economy[3, 41, 43] enables the regulation of supply and demand of resources, offers incentive to the resource owners for leasing, and promotes QoS based resource allocation. The Grid-Federation consists of the cluster owners as resource providers and the end-users as resource consumers. End-users are also likely to be topologically distributed, having different performance goals, objectives, strategies and demand patterns. We focus on optimizing the resource provider's objective and resource consumer's utility functions by using a quoting mechanism. The Grid-Federation consists of cluster resources distributed across multiple organizations and administrative domains. To enable policy based co-ordinated resource sharing between these clusters, we de-fine and model a new RMS system, which we call Grid Federation Agent (GFA). It is a two layer resource management system, managing underlying cluster resources in conjunction with a LRMS. A cluster can become a member of the federation by instantiating a GFA component. GFA acts as a resource co-coordinator in the federated space, spanning over all the clusters. These GFAs in the federation inter-operate using an agreed communication primitive over the shared federation directory.

The model defines two functional units of a GFA: (1) *distributed information manager* (DIM) and (2) *resource manager*. The DIM performs tasks like resource discovery and advertisement through well defined primitives. It interacts with an underlying shared federation directory (shown in Fig.(1)). Recall that, we assume that the directory information is shared using some efficient protocol (e.g. a P2P protocol). In this case the P2P system provides a decentralized database with efficient updates and range query capabilities. Individual GFAs access the directory information using the interface shown in Fig.1, i.e. subscribe, quote, unsubscribe, query. In this paper, we are not concerned with the specifics of the interface (which can be found in[35]). The resource discovery function includes searching for suitable cluster resources while resource advertisement is concerned with advertising resource capability (with pricing policy) to other clusters in the federation. The federation directory maintains quotes or advertised costs from each GFA in the federation. Each quote consists of a resource description $R_i$, for cluster $i$, and a cost $c_i$ for using that resource, configured by respective cluster owners. Using $R_i$ and $c_i$, a GFA can determine the cost of executing a job on cluster $i$ and the time taken, assuming that cluster $i$ has no load. The actual load of the cluster needs to be determined dynamically and the load can lead to changes in time taken (for job completion). In this paper, we assume that $c_i$ remains static throughout the simulations. Each GFA can query the federation directory to find the $k$-th fastest cluster or the $k$-th cheapest cluster. We assume the query process is optimal, i.e. that it takes $O(\log n)$ messages[15] to query the directory, when there are $n$ GFAs in the system. In this paper, we consider the number of additional messages that are used to satisfy our Grid-Federation scheduling process.

The resource manager is responsible for local job superscheduling. Further, it manages the execution of remote jobs in conjunction with the LRMS on the local resource. *Local jobs* refer to the jobs submitted by the local population of users. While *remote jobs* refer to the incoming jobs from remote GFAs. The resource manager provides admission control facility at each in the federation. GFAs undertake one-to-one admission control negotiation with the remote site GFA's resource manager before submitting a job. The admission control negotiation is the enquiry message sent by a remote GFA whether the job can be completed within the specified deadline. Following this, the resource manager queries LRMS about local job queue size, expected job response time and resource utilization status. If the LRMS reports that the job can be completed within the specified deadline, then the admission control acceptance message is sent to the requesting remote GFA. On receiving the acceptance message, remote GFA sends the job.

The proposed Grid-Federation mechanism can leverage services of Grid-Bank [4] for credit management. The participants in the system can use Grid-Bank to exchange Grid Dollars.

## 2.1 General Grid-Federation superscheduling technique

In this section we describe our general Grid-Federation scheduling technique. In Fig.1 a user who is local to GFA 3 is submitting a job. If the user's job QoS can't be satisfied locally then GFA 3 queries the federation directory to obtain the quote of the 1-st fastest or 1-st cheapest cluster. In this case, the federation directory returns the quote advertised by GFA 2. Following this, GFA 3 sends a negotiate message (enquiry about QoS guarantee in terms of response time) to GFA 2. If GFA has too much load and cannot complete the job within the deadline then GFA 3 queries the federation directory for the 2-nd cheapest/fastest GFA and so on. The query-negotiate process is repeated until GFA 3 finds a GFA that can schedule the job (in this example the job is finally scheduled on cluster 4).

Every federation user must express how much he is willing to pay, called a *budget*, and required response time, called a *deadline*, for his job number $j$. In this work, we say that a job's QoS has been satisfied if the job is completed within budget and deadline, otherwise it is not satisfied. Every cluster in the federation has its own resource set $R_i$ which contains the definition of all resources owned by the cluster and ready to be offered. $R_i$ can include information about the CPU architecture, number of processors, RAM size, secondary storage size, operating system type, etc. In this work, $R_i = (p_i, \mu_i, \gamma_i)$ which includes the number of processors, $p_i$, their speed, $\mu_i$ and underlying interconnect network bandwidth $\gamma_i$. We assume that there is always enough RAM and correct operating system conditions, etc. The cluster owner charges $c_i$ per unit time or per unit of million instructions (MI) executed, e.g. per 1000 MI.

We write $J_{i,j,k}$ to represent the $i$-th job from the $j$-th user of the $k$-th resource. A job consists of the number of processors required, $p_{i,j,k}$, the job length, $l_{i,j,k}$ (in terms of instructions), the budget, $b_{i,j,k}$, the deadline or maximum delay, $d_{i,j,k}$ and the communication overhead, $\alpha_{i,j,k}$.

To capture the nature of parallel execution with message passing overhead involved in the real application, we considered a part of total execution time as the communication overhead and remaining as the computational time. In this work, we consider the network communication overhead $\alpha_{i,j,k}$ for a parallel job $J_{i,j,k}$ to be randomly distributed over the processes. In other words, we don't consider the case e.g. when a parallel program written for a hypercube is mapped to a mesh architecture. We assume that the communication overhead parameter $\alpha_{i,j,k}$ would scale the same way over all the clusters depending on $\gamma_i$. The total data transfer involved during a parallel job execution is given by

$$\Gamma(J_{i,j,k}, R_k) = \alpha_{i,j,k}\, \gamma_k \qquad (1)$$

The time for job $J_{i,j,k} = (p_{i,j,k}, l_{i,j,k}, b_{i,j,k}, d_{i,j,k}, \alpha_{i,j,k})$ to execute on resource $R_m$ is

$$D(J_{i,j,k}, R_m) = \frac{l_{i,j,k}}{\mu_m\, p_{i,j,k}} + \frac{\Gamma(J_{i,j,k}, R_k)}{\gamma_m} \qquad (2)$$

$$D(J_{i,j,k}, R_m) = \frac{l_{i,j,k}}{\mu_m\, p_{i,j,k}} + \frac{\alpha_{i,j,k}\, \gamma_k}{\gamma_m} \qquad (3)$$

and the associated cost is

$$B(J_{i,j,k}, R_m) = c_m \frac{l_{i,j,k}}{\mu_m\, p_{i,j,k}}. \qquad (4)$$

If $s_{i,j,k}$ is the time that $J_{i,j,k}$ is submitted to the system then the job must be completed by time $s_{i,j,k} + d_{i,j,k}$.

## 2.2 QoS driven resource allocation algorithm for Grid-Federation

We consider a deadline and budget constrained (DBC) scheduling algorithm, or cost-time optimization scheduling. The federation user can specify any one of the following optimization strategies for their jobs:

- optimization for time (OFT) – give minimum possible response time within the budget limit;
- optimization for cost (OFC) – give minimum possible cost within the deadline.

For each job that arrives at a GFA, called the local GFA, the following is done:

1. Set $r = 1$.

2. If OFT is required for the job then query the federation directory for the $r$-th fastest GFA; otherwise OFC is required and the query is made for the $r$-th cheapest GFA. Refer to the result of the query as the remote GFA.

3. The local GFA sends a message to the remote GFA, requesting a guarantee on the time to complete the job.

4. If the remote GFA confirms the guarantee then the job is sent, otherwise $r := r + 1$ and the process iterates through step 2.

Recall that we assume each query takes $O(\log n)$ messages and hence in this work we use simulation to study how many times the iteration is undertaken, on a per job basis and on a per GFA basis. The remote GFA makes a decision immediately upon receiving a request as to whether it can accept the job or not. If the job's QoS parameters cannot be satisfied (after iterating up to the greatest $r$ such that GFA could feasibly complete the job) then the job is dropped.

Effectively, for job $J_{i,j,k}$ that requires OFC then GFA $m$ with $R_m$ is chosen such that $B(J_{i,j,k}, R_m) = \min_{1 < m' \leq n}\{B(J_{i,j,k}, R_{m'})\}$, and $D(J_{i,j,k}, R_m) \leq s_{i,j,k} + d_{i,j,k}$. Similarly, for OFT then GFA $m$ is chosen such that $D(J_{i,j,k}, R_m) = \min_{1 < m' \leq n}\{D(J_{i,j,k}, R_{m'})\}$, and $B(J_{i,j,k}, R_m) \leq b_{i,j,k}$.

## 2.3 Grid-Federation coordination technique

Currently, the coordination methodology in the Grid-Federation is based on the one-to-one admission control negotiation message. GFAs undertake one-to-one negotiation before submitting a job. The GFA local to the submitted job sends admission control negotiate message to the remote GFA, requesting a guarantee on the total job completion time. If the remote GFA can complete the job within the specified time, then the admission control acceptance message is sent back. Following this, the GFA sends the job. The inter-GFA coordination scheme prevents the GFAs from submitting unlimited amount of jobs to the resources. However, our initial set of experiments do not evaluate the coordination scenario i.e. we don't present the experiments which compares the resource utilization scenario with and without the coordination scheme. Further, the current coordination scheme can be improved by making GFAs dynamically update their local resource utilization metrics into the decentralized federation directory. This can significantly reduce the number of negotiation messages required to schedule a job. We intend to consider these issues and relevant evaluation in our future work.

## 2.4 Quote value

We assume $c_i$ remains static throughout the simulations. In this work, we are only interested in studying the effectiveness of our Grid-Federation superscheduling algo-rithm based on the static access charge $c_i$. Analyzing different pricing algorithm based on supply and demand function is a vast research area. Investigating how the cluster owners determine the price[19, 40, 45] of their commodity is subject of future work. In simulations, we configure $c_i$ using the function:

$$c_i = f(\mu_i) \qquad (5)$$

where,

$$f(\mu_i) = \frac{c}{\mu}\,\mu_i \qquad (6)$$

$c$ is the access price and $\mu$ is the speed of the fastest resource in the Grid-Federation.

## 2.5 User budget and deadline

While our simulations in the next section use trace data for job characteristics, the trace data does not include user specified budgets and deadlines on a per job basis. In this case we are forced to fabricate these quantities and we include the models here.

For a user, $j$, we allow each job from that user to be given a budget (using Eq. 4),

$$b_{i,j,k} = 2\,B(J_{i,j,k}, R_k). \qquad (7)$$

In other words, the total budget of a user over simulation is unbounded and we are interested in computing the budget that is required to schedule all of the jobs.

Also, we let the deadline for job $i$ (using Eq. 2) be

$$d_{i,j,k} = 2\,D(J_{i,j,k}, R_k). \qquad (8)$$

we assign two times the value of total budget and deadline for the given job, as compared to the expected budget spent and response time on the originating resource.

# 3 Experiments and analysis

## 3.1 Workload and resource methodology

We used trace based simulation to evaluate the effectiveness of the proposed system and the QoS provided by the resource allocation algorithm. The workload trace data was obtained from[1]. The trace contains real time workload of various resources/supercomputers that are deployed at the Cornell Theory Center (CTC SP2), Swedish Royal Institute of Technology (KTH SP2), Los Alamos National Lab (LANL CM5), LANL Origin 2000 Cluster (Nirvana) (LANL Origin), NASA Ames (NASA iPSC) and San-Diego Supercomputer Center (SDSC Par96, SDSC Blue, SDSC SP2) (See Table 1). The workload trace is a record of usage data for parallel jobs that were submitted to various resource facilities. Every job arrives, is allocated one or more processors for a period of time, and then leaves the system. Furthermore, every job in the workload has an associated arrival time, indicating when it was submitted to the scheduler for consideration. As the experimental trace data does not include details about the network communication overhead involved for different jobs, we artificially introduced the communication overhead element as 10% of the total parallel job execution time. The simulator was implemented using GridSim[14] toolkit that allows modeling and simulation of distributed system entities for evaluation of scheduling algorithms. To enable parallel workload simulation with GridSim, we extended the existing GridSim's Alloc Policy and Space Shared entities.

Our simulation environment models the following basic entities in addition to existing entities in GridSim:

- local user population – models the workload obtained from trace data;
- GFA – generalized RMS system;
- GFA queue – placeholder for incoming jobs from local user population and the federation;
- GFA shared federation directory – simulates an efficient distributed query process such as peer-to-peer.

For evaluating the QoS driven resource allocation algorithm, we assigned a synthetic QoS specification to each resource including the Quote value (price that a cluster owner charges for service), having varying MIPS rating and underlying network communication bandwidth. The simulation experiments were conducted by utilizing workload trace data over the total period of two days (in simulation units) at all the resources. We consider the following resource sharing environment for our experiments:

- independent resource – Experiment 1;
- federation without economy – Experiment 2;
- federation with economy – Experiments 3, 4 and 5.

## 3.2 Experiment 1 – independent resources

In this experiment the resources were modeled as an independent entity (without federation). All the workload submitted to a resource is processed and executed locally (if possible). In Experiment 1 (and 2) we consider, if the user request can not be served within requested deadline, then it is rejected otherwise it is accepted. During Experiment 1 (and 2), we evaluate the performance of a resource in terms of average resource utilization (amount of real work that a resource does over the simulation period excluding the queue processing and idle time), job acceptance rate (total percentage of jobs accepted) and conversely the job rejection rate (total percentage of jobs rejected). The result of this experiment can be found in Table2. Experiment 1 is essentially the control experiment that is used as a benchmark for examining the affects of using federated(with and without economy) sharing of resources.

Table 1: Workload and Resource Configuration

| Index | Resource / Cluster Name | Trace Date | Processors | MIPS (rating) | Jobs | Quote(Price) | NIC to Network Bandwidth (Gb/Sec) |
|---|---|---|---|---|---|---|---|
| 1 | CTC SP2 | June96-May97 | 512 | 850 | 79,302 | 4.84 | 2 |
| 2 | KTH SP2 | Sep96-Aug97 | 100 | 900 | 28,490 | 5.12 | 1.6 |
| 3 | LANL CM5 | Oct94-Sep96 | 1024 | 700 | 201,387 | 3.98 | 1 |
| 4 | LANL Origin | Nov99-Apr2000 | 2048 | 630 | 121,989 | 3.59 | 1.6 |
| 5 | NASA iPSC | Oct93-Dec93 | 128 | 930 | 42,264 | 5.3 | 4 |
| 6 | SDSC Par96 | Dec95-Dec96 | 416 | 710 | 38,719 | 4.04 | 1 |
| 7 | SDSC Blue | Apr2000-Jan2003 | 1152 | 730 | 250,440 | 4.16 | 2 |
| 8 | SDSC SP2 | Apr98-Apr2000 | 128 | 920 | 73,496 | 5.24 | 4 |

### 3.3 Experiment 2 – with federation

In this experiment, we analyzed the workload processing statistics of various resources when part of the Grid-Federation but not using an economic model. In this case the workload assigned to a resource can be processed locally. In case a local resource is not available then online scheduling is performed that considers the resources in the federation in decreasing order of their computational speed. We also quantify the jobs depending on whether they are processed locally or migrated to the federation. Table3 describes the result of this experiment.

### 3.4 Experiment 3 – with federation and economy

In this experiment, we study the computational economy metaphor in the Grid-Federation. In order to study economy based resource allocation mechanism, it was necessary to fabricate user budgets and job deadlines. As the trace data does not indicate these QoS parameters, so we assigned them using Eqs. 7,8 to all the jobs across the resources. We performed the experiment under 11 different combination of user population profile:
$OFT = i$ and $OFC = 100 - i$ for $i = 0, 10, 20, \ldots, 100$.
Fig.3, 4, 5, 6, 7 and 8 describes the result of this experiment.

### 3.5 Experiment 4 – message complexity with respect to jobs

In this experiment, we consider total incoming and outgoing messages at all GFA's. The various message type includes negotiate, reply, job-submission (messages containing actual job) and job-completion (message containing job output). We quantify the number of local messages (sent from a GFA to undertake a local job scheduling) and remote messages (received at a GFA to schedule a job belonging to a remote GFA in the federation). The experiment was conducted for the same user populations as explained in experiment 3. Fig.9 describes the result of this experiment.

### 3.6 Experiment 5 – message complexity with with respect to system size

This experiment measures the system's performance in terms of the total message complexity involved as the system size grows from 10 to 50. In this case, we consider the average, max and min number of messages (sent/recv) per GFA/per Job basis. Note that, in case $n$ messages are undertaken to schedule a job then it involves traversing (if $n > 2$ then $(n-2)/2$, else $n/2$) entries of the GFA list. To accomplish larger system size, we replicated our existing resources accordingly (shown in Table 1). The experiment was conducted for the same user populations as explained in experiment 3. Fig.10 and 11 describes the result of this experiment. The Java based simulation tool prohibited us from scaling the system further.

### 3.7 Results and observations

#### 3.7.1 Justifying Grid-Federation based resource sharing

During experiment 1 we observed that 5 out of 8 resources remained underutilized (less than 60%). During experiment 2, we observed that overall resource utilization of most of the resources increased as compared to experiment 1 (when they were not part of the federation), for instance resource utilization of CTC SP2 increased from 53.49% to 87.15%. The same trends can be observed for other resources too (refer to Fig.2(a)). There was an interesting observation regarding migration of the jobs between the resources in the federation (load-sharing). This characteristic was evident at all the resources including CTC SP2, KTH SP2, NASA iPSC etc. At CTC, which had total 417 jobs to schedule, we observed that 324 (refer to Table 3 or Fig.2(b)) of them were executed locally while the remaining 93 jobs migrated and executed at some remote resource in the federation. Further, CTC executed 72 remote jobs, which migrated from other resources in the federation.

Table 2: Workload Processing Statistics (Without Federation)

| Index | Resource / Cluster Name | Average Resource Utilization (%) | Total Job | Total Job Accepted(%) | Total Job Rejected(%) |
|---|---|---|---|---|---|
| 1 | CTC SP2 | 53.492 | 417 | 96.642 | 3.357 |
| 2 | KTH SP2 | 50.06438 | 163 | 93.865 | 6.134 |
| 3 | LANL CM5 | 47.103 | 215 | 83.72 | 16.27 |
| 4 | LANL Origin | 44.55013 | 817 | 93.757 | 6.24 |
| 5 | NASA iPSC | 62.347 | 535 | 100 | 0 |
| 6 | SDSC Par96 | 48.17991 | 189 | 98.941 | 1.058 |
| 7 | SDSC Blue | 82.08857 | 215 | 57.67 | 42.3255 |
| 8 | SDSC SP2 | 79.49243 | 111 | 50.45 | 49.54 |

Table 3: Workload Processing Statistics (With Federation)

| Index | Resource / Cluster Name | Average Resource Utilization (%) | Total Job | Total Job Accepted(%) | Total Job Rejected(%) | No. of Jobs Processed Locally | No. of Jobs Migrated to Federation | No. of Remote jobs processed |
|---|---|---|---|---|---|---|---|---|
| 1 | CTC SP2 | 87.15 | 417 | 100 | 0 | 324 | 93 | 72 |
| 2 | KTH SP2 | 68.69 | 163 | 99.38 | 0.61 | 110 | 52 | 35 |
| 3 | LANL CM5 | 67.20 | 215 | 90.69 | 9.30 | 145 | 50 | 70 |
| 4 | LANL Origin | 77.62 | 817 | 98.89 | 1.10 | 733 | 75 | 81 |
| 5 | NASA iPSC | 78.73 | 535 | 99.81 | 0.18 | 428 | 106 | 129 |
| 6 | SDSC Par96 | 79.17 | 189 | 100 | 0 | 143 | 46 | 30 |
| 7 | SDSC Blue | 90.009 | 215 | 98.60 | 1.39 | 105 | 107 | 77 |
| 8 | SDSC SP2 | 87.285 | 111 | 97.29 | 2.70 | 54 | 54 | 89 |

The federation based load-sharing also lead to a decrease in the total job rejection rate, this can be observed in case of resource SDSC Blue where the job rejection rate decreased from 42.32% to 1.39%. Note that, the average job acceptance rate, over all resources in the federation, increased from 90.30% (without federation) to 98.61% (with federation). Thus, for the given job trace, it is preferable to make use of more resources, i.e. to migrate jobs. In other words, the job trace shows the potential for resource sharing to increase utilization of the system.

### 3.7.2 Resource owner perspective

In experiment 3, we measured the computational economy related behavior of the system in terms of its supply-demand pattern, resource owner's incentive (earnings) and end-user's QoS constraint satisfaction (average response time and average budget spent) with varying user population distribution profiles. We study the relationship between resource owner's total incentive and end-user's population profile.

The total incentive earned by different resource owners with varying user population profile can be seen in Fig.3(a). The result shows as expected that the owners (across all the resources) earned more incentive when users sought OFT (Total Incentive $2.30 \times 10^9$ Grid Dollars) (scenario-3) as compared to OFC (Total Incentive $2.12 \times 10^9$ Grid Dollars) (scenario-1). During OFT, we observed that there was a uniform distribution of the jobs across all the resources (refer to Fig.4) and every resource owner earned some incentive. During OFC, we observed a non-uniform distribution of the jobs in the federation (refer to Fig.4). We observed that the resources including CTC SP2, LANL CM5, LANL Origin, SDSC par96 and SDSC Blue earned significant incentives. This can also be observed in their resource utilization statistics (refer to Fig.4). However, the faster resources (e.g. KTH SP2, NASA iPSC and SDSC SP2) remained largely underutilized and did not get significant incentives.

Furthermore, the results indicate an imbalance between the resource supply and demand pattern. As the demand was high for the cost-effective resources compared to the time-effective resources, these time-effective resources remained largely underutilized. In this case, the majority of jobs were scheduled on the cost-effective computational resources (LANL CM5, LANL Origin, SDSC Par96 and SDSC Blue). This is the worst case scenario in terms of resource owner's incentive across all the re-

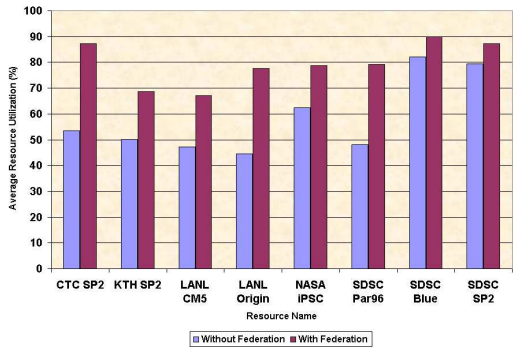

(a) Average resource utilization (%) vs. resource name

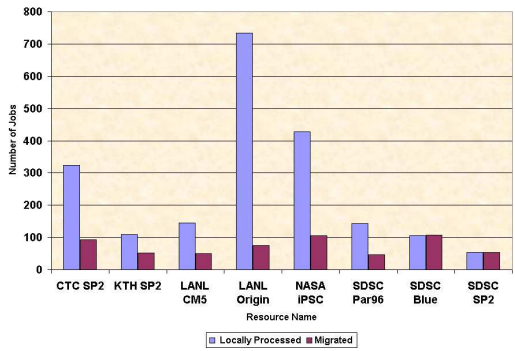

(b) No. of jobs vs. resource name

Figure 2: Resource utilization and job migration plot

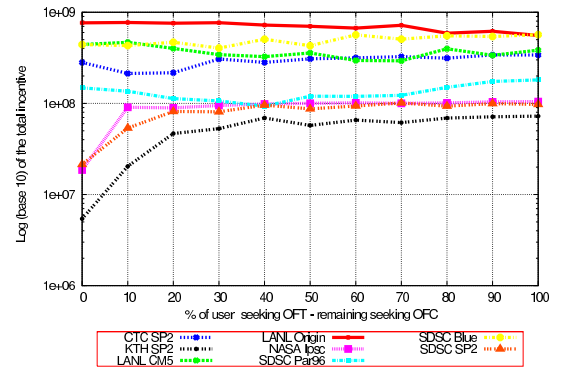

(a) Total incentive (Grid Dollars) vs. user population profile

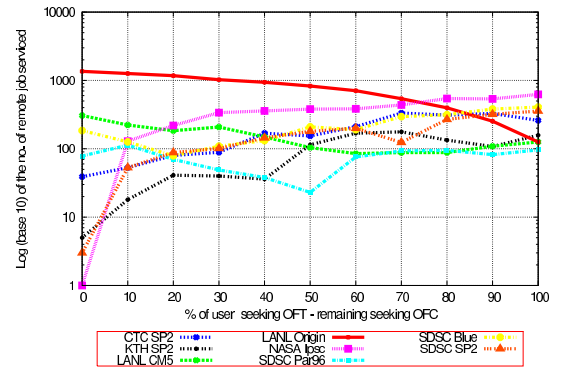

(b) No. of remote job serviced vs. user population profile

Figure 3: Resource owner perspective

sources in the federation. Although, when the majority of end-users sought OFT (more than 50%), we observed uniform distribution of jobs across resources in the federation. Every resource owner across the federation received significant incentive (refer to Fig.3(a)) and had improved resource utilization (refer to Fig.4). These scenarios show balance in the resource supply and demand pattern.

Further, in this case (the majority of users sought OFT (more than 50 percent)), the average resources in terms of cost/time effectiveness (SDSC Par96 and SDSC Blue) made significant incentive (which can also be seen in their average utilization) as compared to when OFC users constituted the majority population. Probably, this is due to computational strength of cost-effective resources (Since LANL Origin and LANL CM5 offered 2048 and 1024 nodes, therefore collectively they satisfied the majority of end-users). So, when OFT users formed the majority it resulted in increased inflow of the remote jobs to these average resources. Similar trends can be identified in their respective total remote job service count (refer to Fig.3(b)). Note that, total remote job service count for cost-effective computational resources (LANL Origin, LANL CM5) decreased considerably as the majority of end-users sought OFT(refer to Fig.3(b)).

Fig.5 shows job migration characteristics at various resources with different population profile. We observed that the most cost-efficient resource (LANL Origin) experienced increased job migration rate in the federation as the majority of its users opted for OFT. Conversely, for the most time-efficient resource (NASA iPSC) we observed slight reduction in the job migration rate.

Thus, we conclude that resource supply (number of resource providers) and demand (number of resource consumers and QoS constraint preference) pattern can determine resource owner's overall incentive and his resource usage scenario.

### 3.7.3 End users perspective

We measured end-users QoS satisfaction in terms of the average response time and the average budget spent under OFC and OFT. We observed that the end-users experienced better average response times (excluding rejected jobs) when they sought OFT for their jobs as compared to OFC (100% users seek OFC) (scenario-1). At LANL Origin (excluding rejected jobs) the average response time for users was $7.865 \times 10^3$ simulation seconds (scenario-1) which reduced to $6.176 \times 10^3$ for OFT (100% users seek OFT) (refer to Fig.7(a)). The end-users spent more budget in the case of OFT as compared OFC (refer to Fig.7(b)). This shows that users get more utility for their QoS constraint parameter response time, if they are willing to spend more budget. Overall, the end-users across all the resources in the federation experi-

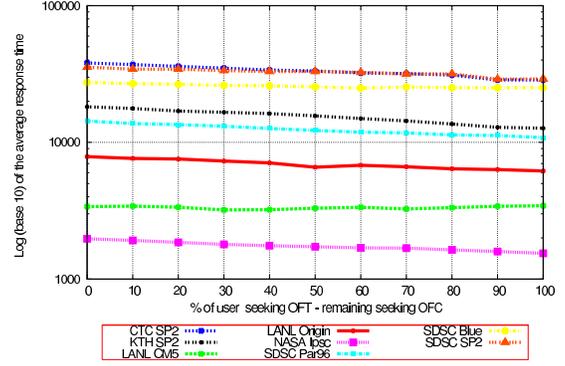

(a) Average response time (Sim Units) vs. user population profile

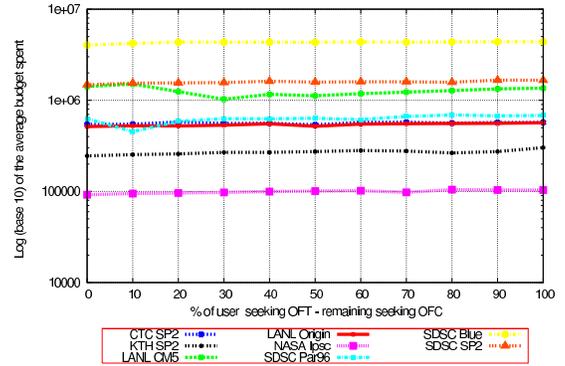

(b) Average budget spent (Grid Dollars) vs. user population profile

Figure 7: Federation user perspective: excluding rejected jobs

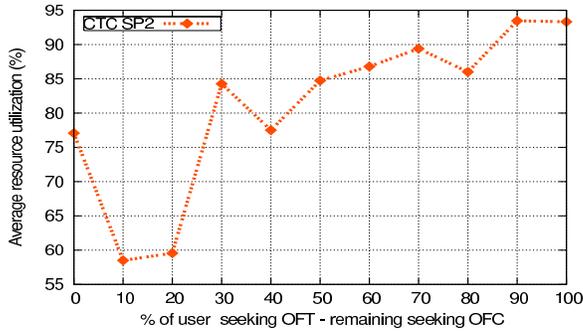
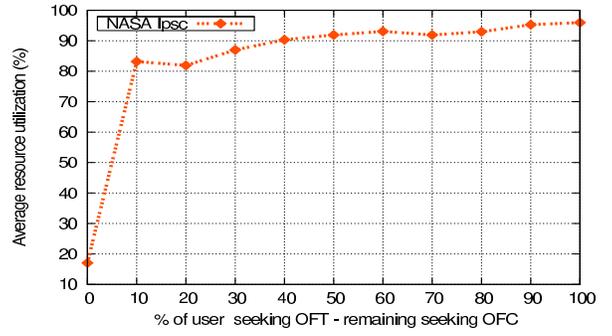
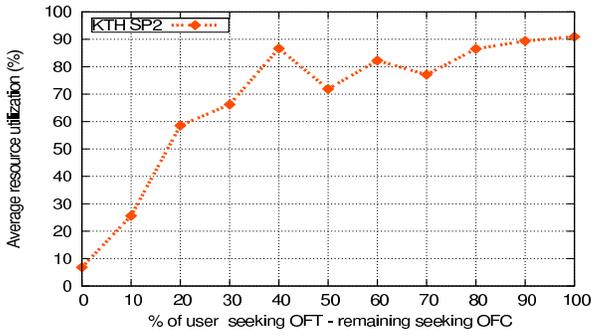
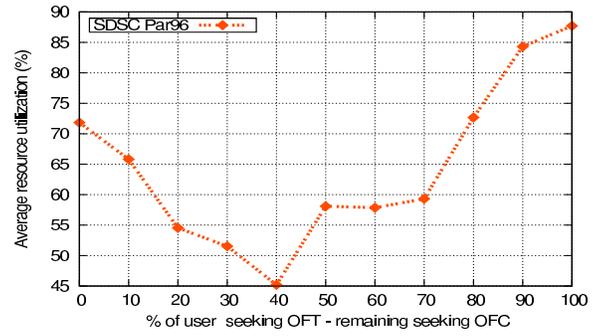
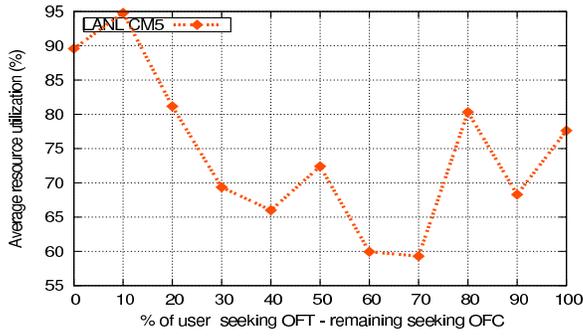
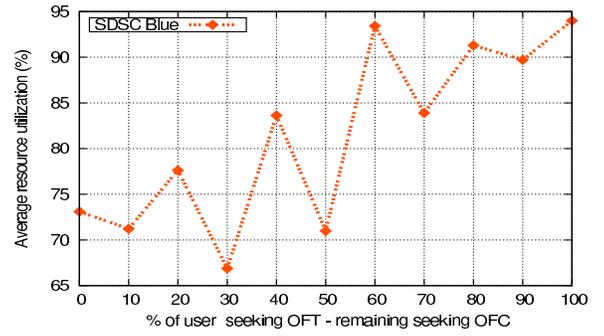
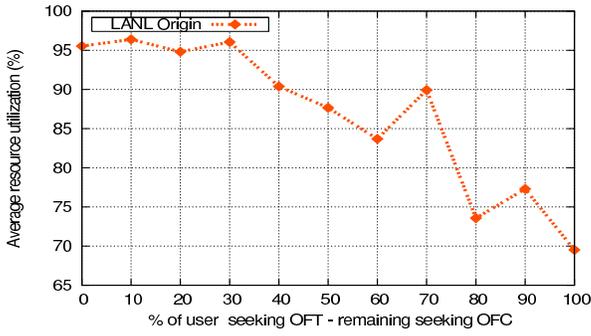
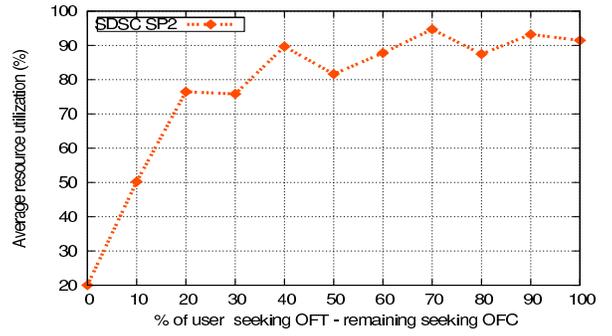

Figure 4: Resource owner perspective: Average resource utilization (%) vs. user population profile

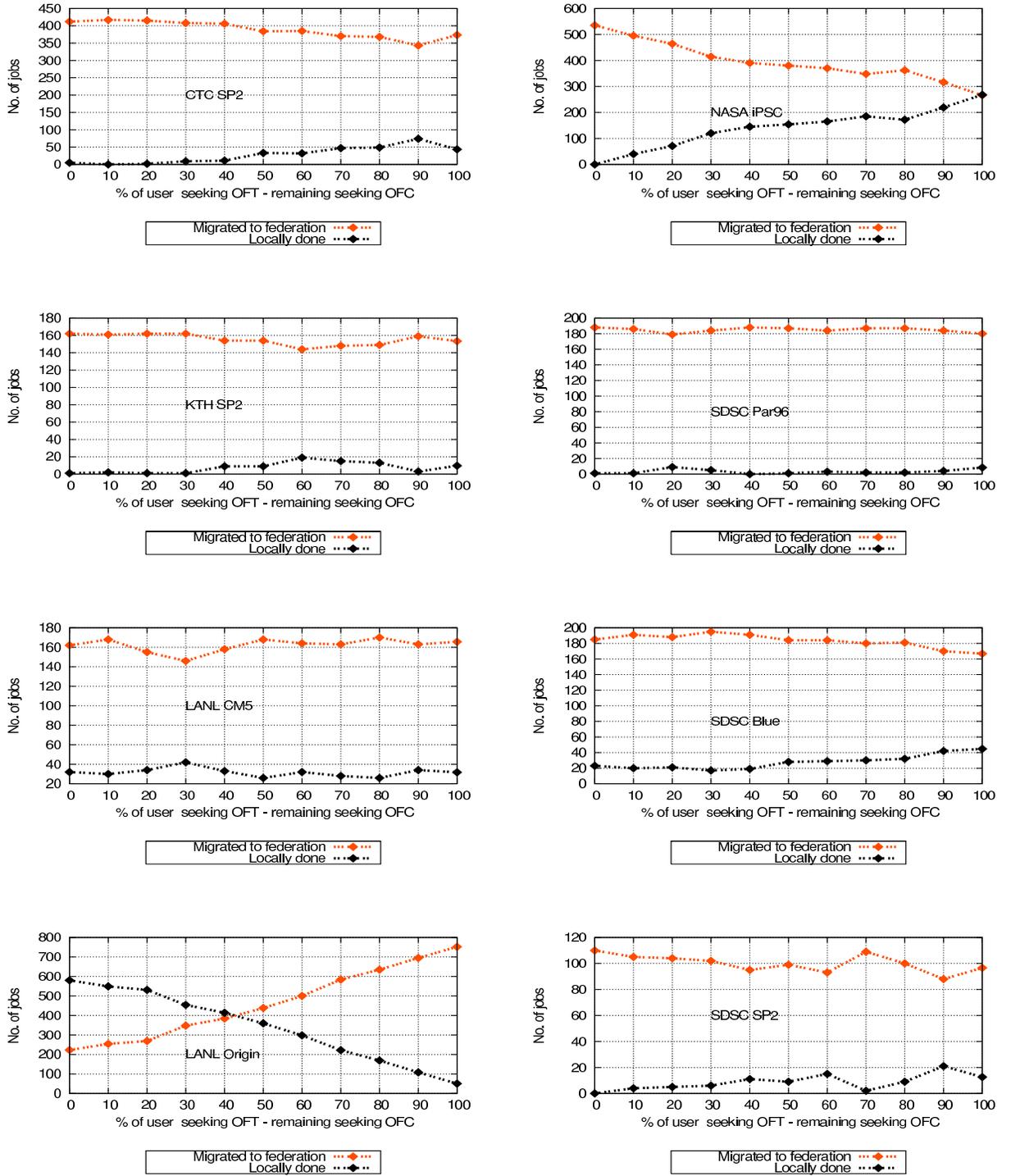

Figure 5: Resource owner perspective: Job processing characteristic vs. user population profile

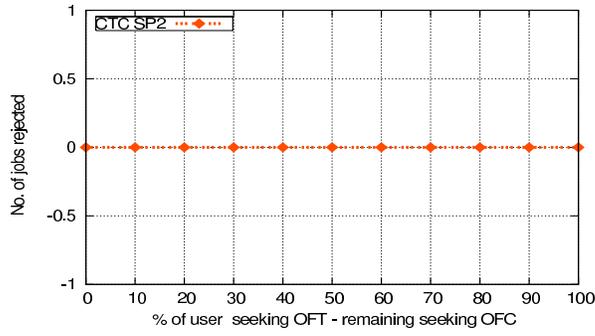
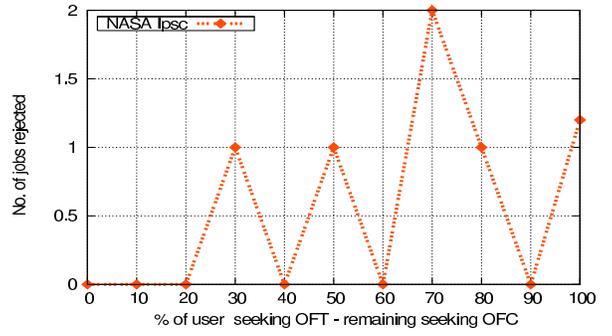
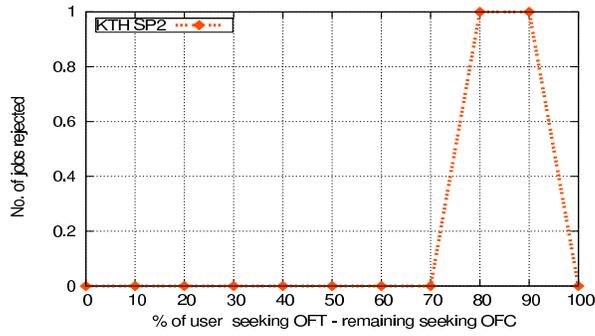
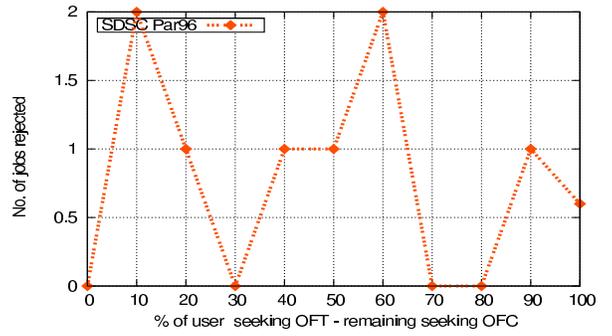
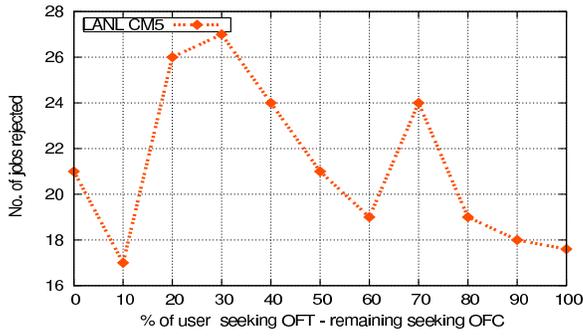
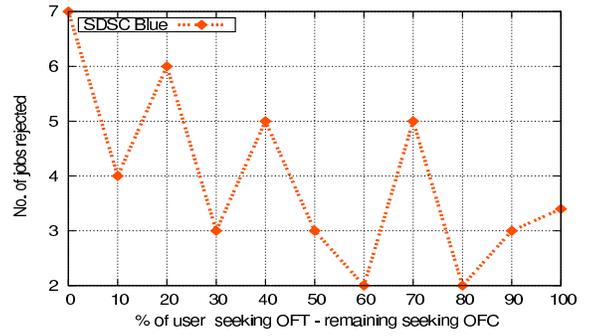
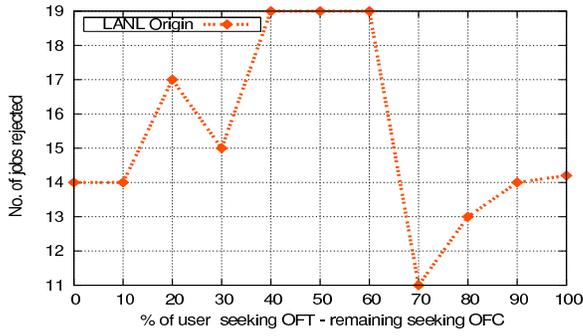
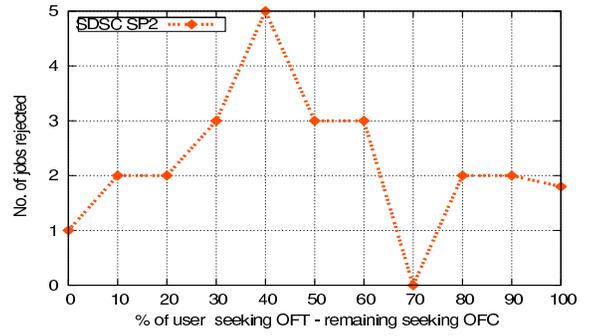

Figure 6: Resource owner perspective: No. of jobs rejected vs. user population profile

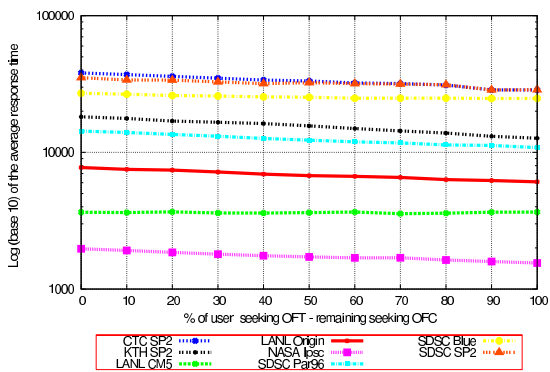

(a) Average response time (Sim Units) vs. user population profile

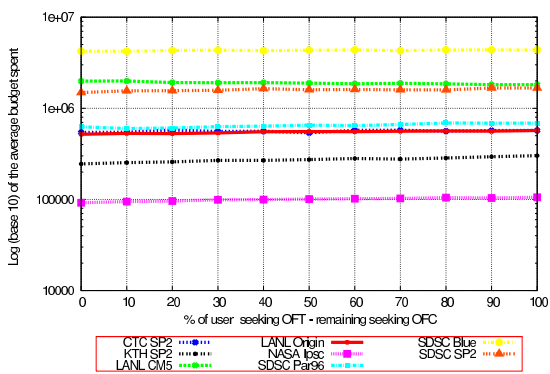

(b) Average budget spent (Grid Dollars) vs. user population profile

Figure 8: Federation user perspective: Including rejected jobs

enced improved response time when the majority constituted OFT population. Although, the end-users belonging to resource LANL CM5 did not had significant change in their response time even with OFT preference. It may be due to their job arrival pattern, that may have inhibited them from being scheduled on the time-efficient resources (though we need to do more investigation including job arrival pattern and service pattern at various resources in order to understand this ).

Note that, Fig.8(a) and Fig.8(b) includes the expected budget spent and response time for the rejected jobs assuming they are executed on the originating resource. Fig.6 depicts the number of jobs rejected across various resources during economy scheduling. During this experiment, we also quantified the average response time and the average budget spent at the fastest (NASA iPSC) and the cheapest resource (LANL Origin) when they are not part of the Grid-Federation (without federation). We observed that the average response time at NASA iPSC was $1.268 \times 10^3$ (without federation) simulation seconds as compared to $1.550 \times 10^3$ (refer to Fig.8(a))) simulation seconds during OFT (100% users seek OFT) (as part of federation). Accordingly, at LANL Origin the average budget spent was $4.851 \times 10^5$ (without federation) Grid Dollars as compared to $5.189 \times 10^5$ (refer to Fig.8(b)) Grid Dollars during OFC (100% users seek OFC) (as part of the federation). Note that, the plots Fig.8(a) and Fig.8(b) do not include the average response time and budget spent for without federation case.

Clearly, this suggests that although federation-based resource sharing leads to better optimization of objective functions for the end-users across all the resources in the federation, sometimes it may be a disadvantage to the users who belong to the most efficient resources (in terms of time or cost).

### 3.7.4 Remote and Local message complexity

In experiment 4, we measured the total number of messages sent and received at various GFA's in the federation with varying user population profiles. Fig.9 shows the plot of the local and remote message count at various GFAs in the federation during economy scheduling. When 100% users seek OFC, we observed that resource LANL Origin received maximum remote messages ($6.407 \times 10^3$ messages) (refer to Fig.9(a)) followed with LANL CM5 (the second cheapest). LANL Origin offers the least cost, so in this case every GFA in the federation attempted to migrate their jobs to LANL Origin, hence leading to increased inflow of the remote messages. While when 100% users seek OFT, we observed maximum number of remote messages at the resource NASA iPSC (refer to Fig.9(a)) followed by SDSC SP2 (the second fastest). Since, these resources were time-efficient, therefore all the GFAs at-

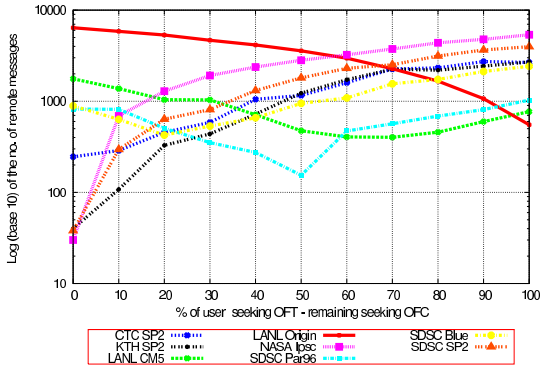

(a) No. of remote messages vs. user population profile

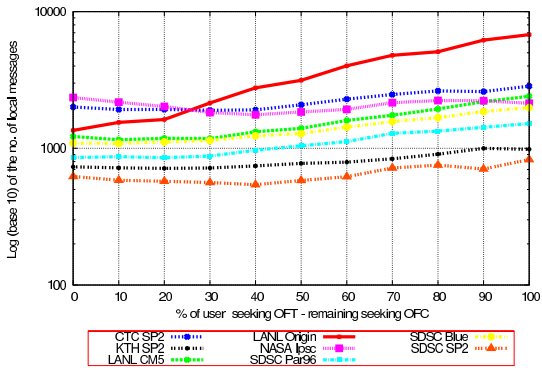

(b) No. of local messages vs. user population profile

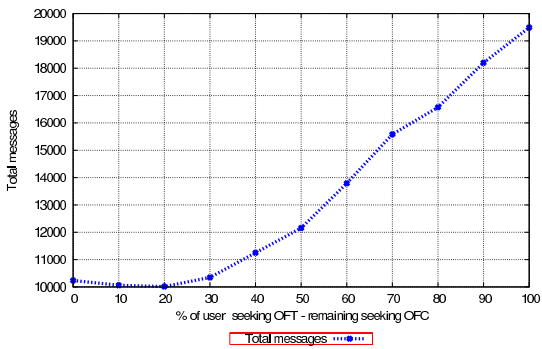

(c) Total messages vs. user population profile

Figure 9: Remote-Local message complexity

tempted to transfer their jobs to them. The total messages involved during this case was $1.948 \times 10^4$ as compared to $1.024 \times 10^4$ during OFC. This happened because the resources LANL Origin and LANL CM5 had 2048 and 1024 computational nodes and a fewer number of negotiation messages were undertaken between GFA's for the job scheduling.

Fig.9(b) shows total number of local messages undertaken at a resource for scheduling work. The results shows, as more users sought OFT, it resulted in increased local message count at cost-effective resources (LANL Origin, LANL CM5). Conversely, faster resources experienced greater remote message count. With 50% seek OFC and 50% seek OFT, we observed uniform distribution of local and remote messages across the federation (refer to Fig.9(a)).

To summarize, we observed linear increase in the total message count with increasing number of the end-users seeking OFT for their jobs (refer to Fig.9(c)). Hence, this suggests that the resource supply and demand pattern directly determines the total number of messages undertaken for the job scheduling in the computational economy based Grid system.

Overall, it can be concluded that the population mix of users in which 70% seek OFC and 30% seek OFT seems most suitable from the system and a resource owner perspective. In this case, we observed uniform distribution of jobs, incentives across the resources. Further, this population mix does not lead to excessive message count as compared to other population mix having greater percentage of users seeking OFT.

### 3.7.5 System's scalability perspective

In experiment 5, we measured the proposed system's scalability with increasing numbers of resource consumers and resource providers. The first part of this experiment is concerned with measuring the average number of messages required to schedule a job in the federation as the system scales. We observed that at a system size of 10, OFC scheduling required an average 5.55 (refer to Fig.10(b)) messages as compared to 10.65 for OFT (Fig.10(b)). As the system scaled to 50 resources, the average message complexity per job increased to 17.38 for OFC as compared to 41.37 during OFT. This suggests that OFC job scheduling required less number of messages than OFT job scheduling, though we need to do more work to determine whether this is due to other factors such as budgets/deadlines assigned to jobs. We also measured the average number of (sent/received) messages at a GFA while scaling the system size (refer to Fig.11). During OFC with 10 resources, a GFA sent/received an average $2.836 \times 10^3$ (refer to Fig.11(b)) messages to undertake scheduling work in the federation as compared to

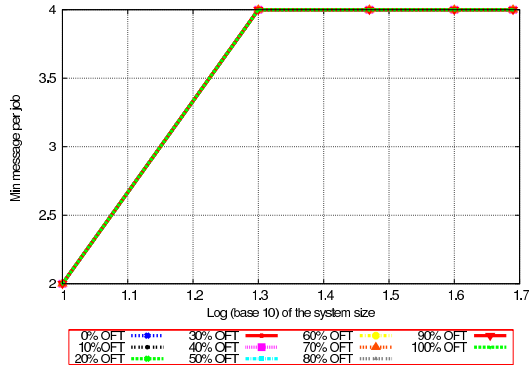

(a) Min message per job vs. system size

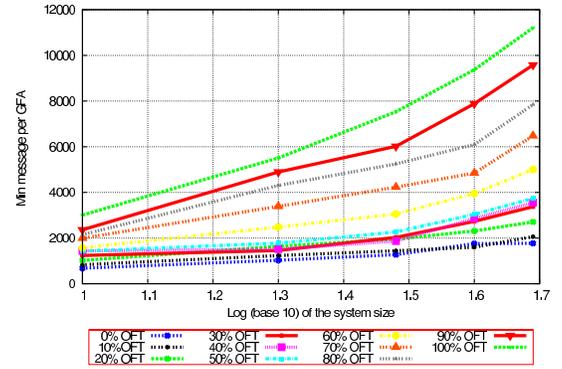

(a) Min message per GFA vs. system size

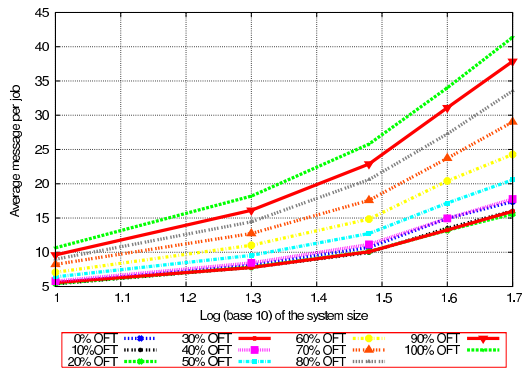

(b) Average message per job vs. system size

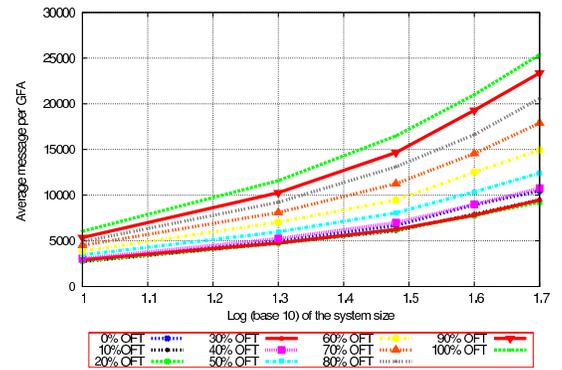

(b) Average message per GFA vs. system size

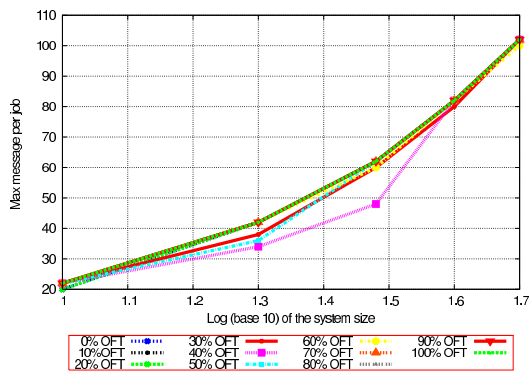

(c) Max message per job vs. system size

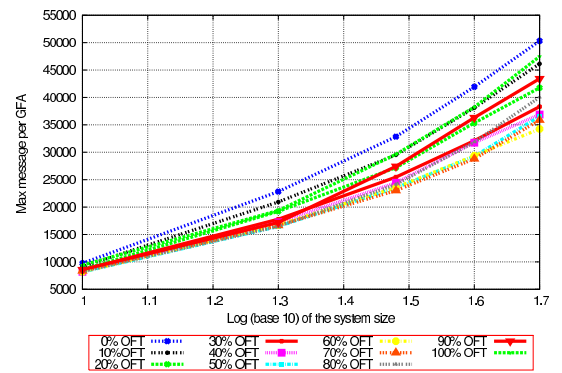

(c) Max message per GFA vs. system size

Figure 10: System's scalability perspective: Message complexity per job with increasing system size

Figure 11: System's scalability perspective: Message complexity per GFA with increasing system size

$6.039 \times 10^3$ (refer to Fig.11(b)) messages during OFT. With 40 resources in the federation, the average message count per GFA increased to $8.943 \times 10^3$ for OFC as regards to $2.099 \times 10^4$ messages for OFT.

Figures 10(b) and 11(b) suggests that the user population including 10%, 20% or 30% OFT seekers involves less number of messages per job/per GFA basis in comparison to 0% OFT seekers. However, with further increase in OFT seekers generates more messages per job/per GFA basis.

From figures 10(b) and 11(b), note that the average message count grows relatively slowly to an exponential growth in the system size. Thus, we can expect that the average message complexity of the system is scalable to a large system size. More analysis is required to understand the message complexity in this case. However, the maximum message count suggests that some parts of the system are not scalable and we need to do more work to avoid these worst cases, e.g. by incorporating more intelligence into the shared federation directory.

Overall, we averaged the budget spent for all the users in the federation during OFC and without federation (independent resources). We observed that during OFC, the average budget spent was $8.874 \times 10^5$ Grid Dollars (we included the expected budget spent of rejected jobs on the originating resource) as compared to $9.359 \times 10^5$ during without federation. However, at most popular resource (LANL Origin) the average budget spent for local users during OFC was $5.189 \times 10^5$ as compared to $4.851 \times 10^5$ during without federation. Similarly, we averaged the response time for all the users in the federation during OFT and without federation. We observed that during OFT, the average response time was $1.171 \times 10^4$ simulation units (we included the expected response time of rejected jobs on the originating resource) as compared to $1.207 \times 10^4$ during without federation. But at the most popular resource (NASA iPSC) the average response time for local users during OFT was $1.550 \times 10^3$ as compared to $1.268 \times 10^3$ during without federation. Clearly, this suggests that while some users that are local to the popular resources can experience higher cost or longer delays during the federation based resource sharing but the overall users' QoS demands across the federation are better met.

Finally, our experiments suggest that the population mix of users in which 70% seek OFC and 30% seek OFT seems most suitable from the system and a resource owner perspective. In this case, we observed uniform distribution of jobs, and incentives across the resources. Further, this population mix does not lead to excessive message count as compared to other population mixes having greater percentage of users seeking OFT.

## 4 Related Work

Resource management and scheduling for parallel and distributed systems has been investigated extensively in the recent past (Apples, NetSolve[16], Condor, LSF, SGE, Punch, Legion, Condor-Flock, NASA-Superscheduler, Nimrod-G and Condor-G). In this paper, we mainly focus on superscheduling systems that allow scheduling jobs across wide area distributed clusters. We highlight the current scheduling methodology followed by Grid superscheduling systems including NASA-superscheduler, Condor-Flock(based on P2P substrate Pastry[36]), Legion-based federation and Resource Brokers. Furthermore, we also discuss some computational economy based cluster and grid systems.

The work in[38] models a grid superscheduler architecture and presents three different distributed job migration algorithms. They consider job scheduling in computational grids through autonomous local schedulers that cooperate through a superscheduler (grid scheduler) using grid middleware. Each resource is modeled to have a grid scheduler (GS), grid middleware(GM) and a local scheduler (LRMS). In the distributed setting, every GS has affinity with its LRMS. The GS is responsible for resource discovery, monitoring system status (utilization, network condition), coordinating job migration related information with other GS in the system. The GS manages the grid queue which is a initial placeholder for all incoming jobs at a resource. Incoming job types include the local jobs and remote jobs. The local jobs are submitted by local user population while the remote jobs are migrated by other GSes in the system. While the LRMS manages the local queue which is a placeholder for migrated remote jobs and set of local jobs allocated by GS to the underlying resource. Whenever, a job is submitted to the grid queue, the GS queries its LRMS though GM for expected average wait time(AWT) in the local queue. If the AWT is below the predetermined threshold value $\phi$(driven by local site sharing policy) then the job is moved to the local queue. However, if the AWT exceeds $\phi$ then one of the three distributed job-migration mechanism is initiated by the GS. The approaches differ in the way communication is carried out between the various GSes in order to facilitate the load-balancing. These job-migration algorithms are referred to as (i) Sender-Initiated(S-I); (ii) Receiver-Initiated(RI); and (iii) Symmetrically-Initiated(Sy-I). In S-I, the GS sends a resource demand query for a job to all other GSes in the system through its GM. So, effectively this approach is based on one-to-all broadcast communication mechanism. In response to a GS resource query, every GS sends back the expected AWT, expected run time(ERT) for the requested job and current resource utilization status(RUS). The value for the parameters AWT, ERT and RUS is obtained by consulting the respective

LRMS. Based on the responses, the initiator GS computes the potential turnaround cost(TC) for every candidate GS. TC is computed as the sum of AWT and ERT. The GS with minimum TC is chosen for job-migration. In case two GS have the same value for TC, then RUS is utilized as a tie-breaker. Hence, the resource which can give least response time for the job is chosen. Under R-I job-migration approach, every GS periodically checks its own RUS at time interval $\sigma$. if the RUS is below a certain pre-defined threshold $\delta$ then the GS volunteers itself for job-migration. It broadcasts its RUS parameter to all GSes in the system. In case, a GS needs to migrate its local job then it initiates S-I base job migration with the volunteer nodes. Finally, the Sy-I approach works in both active and passive mode. Under this approach both S-I and R-I based job migration algorithm can be initiated by the GSes in the system. Effectively, the job scheduling is based on broadcast communication approach that may generate a large number of network messages. Such scheduling approach has serious scalability concerns. In contrast to this superscheduling system, our approach differs in the following (i) the job-migration or the load-balancing in the Grid-Federation is driven by user specified QoS constraints and resource owners' sharing policies; (ii) our approach gives a resource owner complete autonomy over resource allocation decision; and (iii) our superscheduling mechanism utilizes decentralized shared federation directory for indexing and querying the resources.

The work in[12] presents a superscheduling system that consists of Internet-wide Condor work pools. They utilize Pastry routing substrate to organize and index the Condor work pool. The resource discovery in previous versions of Condor flock[22] was based on static knowledge and required manual configuration. To an extent, the previous approach was centralized in nature. Pastry arranges the pools on a logical ring (the P2P overlay's node identifier name space) and allows a Condor pool to dynamically join the existing flock structure using the bootstrap node. Activities related to P2P overlay organization and management is carried out by a central work pool manager. However, the superscheduling scheme can only schedule the jobs to the work pools whose node-id is indexed by the local pool managers' routing tables. In other words, the superscheduling decision is based on partial-set of resources and hence it inhibits the system from approach optimal load balancing. Further, broadcast mechanism(sending inquiry message to every work pool in the routing table about resource availability and their willingness to accept jobs) is used to inquire about resource status. Such approach can be very costly in terms of network communication overhead. The superscheduling scheme periodically compares the metrics such as queue lengths, average pool utilization and resource availability scenario, and based on these statistics a sorted list of pools from most suitable to least suitable is formulated. Using this list, a local Condor pool manager chooses appropriate pools for flocking. The superscheduling mechanism is based on system-centric parameters. In contrast, Grid-Federation is based on decentralized shared federation directory, hence our superscheduling mechanism is based on the complete resource set. Further, the superscheduling scheme considers user-centric parameters for job scheduling across the federation.

OurGrid[7] provides a grid superscheduling middleware infrastructure based on the P2P network paradigm. The OurGrid community is basically a collection of a number of OurGrid Peer(OG Peer) that communicate using P2P protocols. Each OG Peer represents a site. Similar to the definition of a P2P system, each site has resource provider as well as resource consumer population. A resource consumer(user) runs a brokering system called MyBroker(a application-level scheduler). Every MyBroker connects to the OurGrid community through its local OG Peer. A resource provider runs the software system called Swan, that facilitates access to his resource for any user in the OurGrid community. The resource sharing in OurGrid is based on P2P file-sharing model such that every participant contributes as well as consumes resources to/from the community. To negate free-riding in a computational grid environment the model defines a new trust and reputation management scheme called *Network of Favors*[6]. Network of Favors promotes load sharing between collaborating sites in the OurGrid, while discouraging the free riders. Further, it maintains one-to-one resource sharing credit between the resource providers. A user submits his application to his MyBroker. Depending on the users' application requirement, MyBroker sends its request for grid machines to the OG Peer. If the machines at local site does not match applications' resource requirement then the request is forwarded to other OG Peers(broadcast) in the community. Depending on the resource availability pattern and initiator sites' reputation, the OG Peers reply to the resource query. In other words, superscheduling in OurGrid is primarily driven by the site's reputation in the community. In contrast, we propose more generalized resource sharing system based on real-market models. Further, our superscheduling system focuses on optimizing resource owners and consumers objective functions.

MOSIX is a cluster management system that applies process migration to enable a loosely coupled Linux cluster to work like a shared memory parallel(SMP) computer. Recently, it has been extended to support a grid of Linux clusters to form a federation[9]. MOSIX Federation(MFED) couples computational clusters under same administrative domain. A basic feature of the federated environment includes automatic load balancing among participant clusters (owned by different de-partments) while preserving the complete site autonomy. Clusters are arranged in hierarchy to form MFED environment. A hierarchical information dissemination scheme, that enables each node to be aware of the latest system wide state. The resource information in system is updated using the randomized gossip algorithm, that requires each node (a machine) to regularly monitor (specified time interval) the state of its resources (CPU usage, current load, memory status, network status) and send this information to a randomly chosen node in the same cluster. Further, this information is exchanged among different clusters at a rate which is proportional to the relative network proximity of clusters. This dynamic resource information is used for inter-cluster and intra-cluster process migration. In other words, the superscheduling decision in MFED driven by load conditions of clusters(system centric parameters). In case, a cluster is found to be heavily loaded then some processes are migrated to other lightly loaded ones. Other key feature of MFED includes supporting dynamic, grid-wide preemptive process migration. Each user in MFED is allowed to create his processes on the nodes belonging to their partition. However, to support dynamic load balancing, a cluster owner can make two sets of machines one for home users while other for remote users. Thus, this allows a resource owner to clearly define what is shared and what is not. In additions to this, the system enforces a process precedence scheme, in which process with higher precedence may push out all the processes with lower precedence (forced preemption). Such precedence is specified by respective owners of the nodes. Other features include flood control which limits the number of remote processes that can be run on a node. Further, processes of a user that may overload a node are not allowed to migrate. In contrast, we propose a more generalized superscheduling system where load-balancing is motivated by resource owners and resource consumers' objective functions. Our system considers scheduling jobs across computational clusters belonging to different administrative domains. Further, we apply the P2P network model to manage resource information thus negating obvious disadvantages of hierarchical approach.

Bellagio[8] is a market-based resource allocation system for federated distributed computing infrastructures. Resource allocation in this system is based on bid-based proportional resource sharing model. Bids for resources are cleared by a centralized auctioneer. Users'(i.e. application superschedulers) discover resources by querying the SWORD[34] system. SWORD is a decentralized resource discovery service that supports multi-attribute queries. SWORD supports queries including per-node characteristics such as load, physical memory, disk space and inter-node network connectivity attributes such as network latency. A bid for resource includes sets of resources desired, processing duration, and the amount of virtual currency which a user is willing to spend. The Centralized auctioneer clears the bid every hour. The resource exchange in the current system is done through virtual currency. Virtual currency is the amount of credit a site has, which is directly determined by the site's overall resource contribution to the federated system. The centralized auctioneer uses the SHARE[21] framework for resolving bids. SHARE allocates resources by clearing a combinatorial auction. In contrast, we propose a decentralized superscheduling system based on commodities markets. Resource allocation decision in our proposed system is controlled by the concerned site , hence providing complete site autonomy.

Tycoon[31] is a distributed market-based resource allocation system. Application scheduling and resource allocation in Tycoon is based on decentralized isolated auction mechanism. Every resource owner in the system runs its own auction for his local resources. Furthermore, auctions are held independently, thus clearly lacking any coordination. Tycoon system relies on centralized Service Location Services(SLS) for index resource auctioneers' information. Auctioneers register their status with the SLS every 30 seconds. In case, a auctioneer fails to update its information within 120 seconds then SLS deletes its entry. Note that, in distributed setting such centralized indexing services can prove to be serious bottleneck in performance and reliability. Application level superschedulers contact the SLS to gather information about various auctioneers in the system. Once this information is available, the superschedulers(on behalf of users) issue bids for different resources(controlled by different auctions) constraint to resource requirement and available budget. In this setting, various superschedulers might end up bidding for small subset of resources while leaving other underutilized. In other words, superscheduling mechanism clearly lacks coordination. A resource bid is defined by the tuple$(h, r, b, t)$ where $h$ is the host to bid on, $r$ is the resource type, $b$ is the number of credits to bid, and $t$ is the time interval over which to bid. Auctioneers determine the outcome by using bid-based proportional resource sharing economy model. In contrast, we propose a mechanism for cooperative and coordinated sharing of distributed clusters based on computational economy. We apply commodity market model for regulating supply and demand of resources in the Grid-Federation.

Legion is an object-based meta-system developed at the University of Virginia. Legion provides a platform to couple heterogeneous, geographically distributed resources. The work[44] proposes federated model for scheduling in wide-area systems and its possible implementation in Legion. The proposed model is based on local schedulers and wide-area schedulers. The wide-area scheduler consults the local site schedulers to obtain candidate machine schedules. The inherent scheduling mechanism is system-centric. Our proposed system applies market based economy principles for resource allocation in the federated environment.

Nimrod-G[3] is an resource management system(RMS) that serves as a resource broker and supports deadline and budget constrained algorithms for scheduling task-farming applications on the platform. It allows the users to lease and aggregate resources depending on their availability, capability, performance, cost, and users QoS constraints. Application scheduling is based on user-centric parameters. The superscheduling mechanism inside the Nimrod-G does not take into account other brokering systems currently present in the system. This can lead to over-utilization of some resources while under-utilization of others. To overcome this, we propose a set of distributed brokers having a transparent co-ordination mechanism.

Libra[39] is a computational economy based cluster-level application scheduler. This system demonstrates that the heuristic economic and QoS driven cluster resource allocation is feasible since it delivers better utility than traditional a system-centric one for the independent job model. Existing versions of Libra lack support for scheduling jobs composed of parametric and parallel models, and a federated resource sharing environment.

REXEC[20] is remote execution environment for a campus-wide network of workstations, which is part of Berkeley Millennium Project. At a command line, the user can specify the maximum credits per minute that he is willing to pay for CPU time. The REXEC client selects a node that fits the user requirements. REXEC allocates resources to user jobs proportional to the user demands. It offers a generic user interface for computational economy on clusters but not a large scale scheduling system. It allocates resources to user jobs proportional to the user valuation irrespective of their job needs, so it is more user centric type. It is targeted towards cluster resource management while in contrast we propose a more generalized grid system.

Finally in Table4, we summarize various superscheduling systems based on underlying network model, scheduling parameter and scheduling mechanism.

## 5 Conclusion

We proposed a new computational economy based distributed cluster resource management system called Grid-Federation. The federation uses agents that maintain and access a shared federation directory of resource information. A cost-time scheduling algorithm was applied to simulate the scheduling of jobs using iterative queries to the federation directory. Our results show that, while the users from popular (fast/cheap) resources have increased competition and therefore a harder time to satisfy their QoS demands, in general the system provides an increased ability to satisfy QoS demands over all users. The result of the QoS based resource allocation algorithm indicates that the resource supply and demand pattern affects resource provider's overall incentive. Clearly, if all users are seeking either time/cost optimization then the slowest/most expensive resource owners will not benefit as much. However if there is a mix of users, some seeking time and some seeking cost optimization then all resource providers gain some benefit from the federation. In our future work we will study to what extent the user profile can change and how pricing polices for resources leads to varied utility of the system. We will also study how the shared federation directory can be dynamically updated with these pricing policies which can lead to co-ordinated QoS scheduling.

We analyzed how the resource supply and demand pattern affects the system scalability/performance in terms of total message complexity. In general, the cost-time scheduling heuristic does not lead to excessive messages, i.e. to excessive directory accesses and we expect the system to be scalable. However it is clear that popular resources can become bottlenecks in the system and so we intend to research ways to avoid such bottlenecking behavior, principally by using coordination via the shared federation directory. Overall, the proposed Grid-Federation, in conjunction with a scalable, shared, federation directory, is a favourable model for building large scale grid systems.

## References

[1] *http://www.cs.huji.ac.il/labs/parallel*.

[2] *http://www.platform.com/products/wm/LSF*.

[3] D. Abramson, R. Buyya, and J. Giddy. A computational economy for grid computing and its implementation in the Nimrod-G resource broker. *Future Generation Computer Systems (FGCS) Journal, Volume 18, Issue 8, Pages: 1061-1074, Elsevier Science, The Netherlands, October*, 2002.

[4] B. Alexander and R. Buyya. Gridbank: A grid accounting services architecture for distributed systems sharing and integration. *Workshop on Internet Computing and E-Commerce, Proceedings of the 17th Annual International Parallel and Distributed Processing Symposium (IPDPS 2003), IEEE Computer Society Press, USA, April 22-26 Nice, France*, 2003.

[5] A. O. Allen. *Probability, Statistics and Queuing Theory with computer science applications*. Academic Press, INC., 1978.

[6] N Andrade, F Brasileiro, W Cirne, and M Mowbray. Discouraging free riding in a peer-to-peer cpu-sharing grid.

Table 4: Superscheduling Technique Comparison

| Index | System Name | Network Model | Scheduling Parameters | Scheduling Mechanism |
|---|---|---|---|---|
| 1 | NASA-Superscheduler | Random | System-centric | Partially coordinated |
| 2 | Condor-Flock P2P | P2P(Pastry) | System-centric | Partially coordinated |
| 3 | Grid-Federation | P2P(Decentralized directory) | User-centric | Coordinated |
| 4 | Legion-Federation | Random | System-centric | Coordinated |
| 5 | Nimrod-G | Centralized | User-centric | Non-coordinated |
| 6 | Condor-G | Centralized | System-centric | Non-coordinated |
| 7 | Our-Grid | P2P | System-centric | Coordinated |
| 8 | Tycoon | Centralized | User-centric | Non-coordinated |
| 9 | Bellagio | Centralized | User-centric | Coordinated |
| 10 | Mosix-Grid | Hierarchical | System-centric | Coordinated |

In *the 13th IEEE International Symposium on High Performance Distributed Computing (HPDC-13)*. IEEE Computer Society,, 2004.

[7] N. Andrade, W. Cirne, F. Brasileiro, and P. Roisenberg. OurGrid: An approach to easily assemble grids with equitable resource sharing. In *Proceedings of the 9th Workshop on Job Scheduling Strategies*. Lecture Notes in Computer Science, 2003.

[8] A. Auyoung, B. Chun, A. Snoeren, and A. Vahdat. Resource allocation in federated distributed computing infrastructures. In *OASIS '04: 1st Workshop on Operating System and Architectural Support for the Ondemand IT InfraStructure, Boston, MA, October*, 2004.

[9] A. Barak, A. Shiloh, and L. Amar. An organizational grid of federated mosix clusters. *Proceedings of the 5th IEEE/ACM International Symposium on Cluster Computing and the Grid (CCGRID'05)*, 2005.

[10] F. Berman and R. Wolski. The apples project: A status report. *Proceedings of the 8th NEC Research Symposium, Berlin, Germany*, 1997.

[11] B. Bode, D. Halstead, R. Kendall, and D. Jackson. PBS: The portable batch scheduler and the maui scheduler on linux clusters. *Proceedings of the 4th Linux Showcase and Conference, Atlanta, GA, USENIX Press, Berkley, CA, October*, 2000.

[12] A. Raza Butt, R. Zhang, and Y. C. Hu. A self-organizng flock of condors. In *SC '03: Proceedings of the 2003 ACM/IEEE conference on Supercomputing*, Washington, DC, USA, 2003. IEEE Computer Society.

[13] R. Buyya, D. Abramson, J. Giddy, and H. Stockinger. Economic models for resource management and scheduling in grid computing. *Special Issue on Grid computing Environment, The Journal of concurrency and Computation:Practice and Experience (CCPE), Volume 14, Issue 13-15, Wiley Press*, 2002.

[14] R. Buyya and M. Murched. Gridsim: A toolkit for the modeling and simulation of distributed resource management and scheduling for grid computing. *Journal of Concurrency and Computation: Practice and Experience;14(13-15), Pages:1175-1220*, 2002.

[15] M. Cai, M. Frank, J. Chen, and P. Szekely. Maan: A Multi-atribute addressable network for grid information services. *Proceedings of the Fourth IEEE/ACM International workshop on Grid Computing;Page(s):184 - 191*, 2003.

[16] H. Casanova and J. Dongara. Netsolve: A network server solving computational science problem. *International Journal of Supercomputing Applications and High Performance Computing;11(3); Pages:212-223*, 1997.

[17] S. Chapin, J. Karpovich, and A. Grimshaw. The legion resource management system. *Proceedings of the 5th Workshop on Job Scheduling Strategies for Parallel Processing, San Juan, Puerto Rico, 16 April, Springer:Berlin*, 1999.

[18] J. Chase, L. Grit, D. Irwin, J. Moore, and S. Sprenkle. Dynamic virtual clusters in a grid site manager. *In the Twelfth International Symposium on High Performance Distributed Computing (HPDC-12), June*, 2003.

[19] J. Q. Cheng and M. P. Wellman. The WALRAS Algorithm: A convergent distributed implementation of general equilibrium outcomes. In *Computational Economics, Volume 12, Issue 1*, pages 1 – 24, Aug 1998.

[20] B. Chun and D. Culler. A decentralized, secure remote execution environment for clusters. *Proceedings of the 4th Workshop on Communication, Architecture and Applications for Network-based Parallel Computing, Toulouse, France*, 2000.

[21] B. Chun, C. Ng, J. Albrecht, D. Parkes, and A. Vahdat. SHARE: Computational resource exchanges for distributed resource allocation. 2004.

[22] D.H.J. Epema, M. Livny, R. van Dantzig, X. Evers, and J. Pruyne. A worldwide flock of condors: Load sharing among workstation clusters. *Future Generation Computer Systems, Vol. 12*, 1996.

[23] I. Foster and C. Kesselman. The grid: Blueprint for a new computing infrastructure. *Morgan Kaufmann Publishers, USA*, 1998.

[24] I. Foster, C. Kesselman, and S. Tuecke. The anatomy of the grid: Enabling scalable virtual organizations. *International Journal of Supercomputer Applications, Vol. 15, No.3*, 2001.

[25] J. Frey, T. Tannenbaum, M. Livny, I. Foster, and S. Tuecke. Condor-G: A computation management agent for multi-institutional grids. In *10th IEEE International Symposium on High Performance Distributed Computing (HPDC-10 '01), 2001*, pages 237 – 246, Washington, DC, USA, 2001. IEEE Computer Society.

[26] W. Gentzsh. Sun grid engine: Towards creating a compute power grid. *Proceedings of the First IEEE/ACM International Symposium on Cluster Computing and the Grid*, 2002.

[27] A. Iamnitchi and I. Foster. On fully decentralized resource discovery in grid environments. *International Workshop on Grid Computing, Denver, CO*, 2001.

[28] IEEE. Ieee std 802.3. Technical report, IEEE, 2002.

[29] J. In, P. Avery, R. Cavanaugh, and S. Ranka. Policy based scheduling for simple quality of service in grid computing. *Proceedings of the 18th International Parallel and Distributed Processing Symposium (IPDPS'04)*, 2004.

[30] N. Kapadia and J. Fortes. Punch: An architecture for web-enabled wide-area network computing. *Cluster computing: The Journal of Networks, Software Tools and Applications;2(2) Pages:153-164*, 1999.

[31] K. Lai, B. A. Huberman, and L. Fine. Tycoon: A distributed market-based resource allocation system. *Technical Report, HP Labs*, 2004.

[32] J. Litzkow, M. Livny, and M. W. Mukta. Condor- a hunter of idle workstations. *IEEE*, 1988.

[33] D. Moore and J. Hebeler. *Peer-to-Peer:Building Secure, Scalable, and Manageable Networks*. McGraw-Hill Osborne, 2001.

[34] D Oppenheimer, J Albrecht, A Vahdat, and D Patterson. Design and implementation tradeoffs for wide-area resource discovery. In *Proceedings of 14th IEEE Symposium on High Performance, Research Triangle Park, NC*, Washington, DC, USA, July 2005. IEEE Computer Society.

[35] R. Ranjan, A. Harwood, and R. Buyya. Grid-federation: A resource management model for cooperative federation of distributed clusters. *Technical Report, GRIDS-TR-2004-10, Grid Computing and Distributed Systems Laboratory, University of Melbourne, Australia*, 2004.

[36] A. Rowstron and P. Druschel. Pastry: Scalable, decentralized object location, and routing for large-scale peer-to-peer systems. In *Middleware'01: Proceedings of IFIP/ACM International Conference on Distributed Systems Platforms*, pages 329–359, Heidelberg, Germany, 2001.

[37] J.M. Schopf. Ten actions when superscheduling. In *Global Grid Forum*, 2001.

[38] H. Shan, L. Oliker, and R. Biswas. Job superscheduler architecture and performance in computational grid environments. In *SC '03: Proceedings of the 2003 ACM/IEEE conference on Supercomputing*, page 44, Washington, DC, USA, 2003. IEEE Computer Society.

[39] J. Sherwani, N. ALi, N. Lotia, Z. Hayat, and R. Buyya. Libra: An economy driven job scheduling system for clusters. *Proceedings of 6th International Conference on High Performance Computing in Asia-Pacific Region (HPC Asia'02)*, 2002.

[40] S. Smale. Convergent process of price adjustment and global newton methods. In *American Economic Review, 66(2)*, pages 284–294, May 1976.

[41] M. Stonebraker, R. Devine, M. Kornacker, W. Litwin, A. Pfeffer, A. Sah, and C. Staelin. An economic paradigm for query processing and data migration in maiposa. *Proceedings of 3rd International Conference on Parallel and Distributed Information Systems, Austin, TX, USA, September 28-30, IEEE CS Press*, 1994.

[42] S. Venugopal, R. Buyya, and L. Winton. A grid service broker for scheduling distributed data-oriented applications on global grids. In *Proceedings of the 2nd workshop on Middleware for grid computing*, pages 75–80, New York, NY, USA, 2004. ACM Press.

[43] C. Waldspurger, T. Hogg, B. Huberman, J. Kephart, and W. Stornetta. Spawn: A distributed computational economy. *IEEE Transactions on Software Engineering , Vol. 18, No.2, IEEE CS Press, USA, February*, 1992.

[44] J. B. Weissman and A. Grimshaw. Federated model for scheduling in wide-area systems. *Proceedings of the Fifth IEEE International Symposium on High Performance Distributed Computing (HPDC), Pages:542-550, August*, 1996.

[45] R. Wolski, J. S. Plank, J. Brevik, and T. Bryan. G-commerce: Market formulations controlling resource allocation on the computational grid. In *IPDPS '01: Proceedings of the 15th International Parallel & Distributed Processing Symposium*, page 46, Washington, DC, USA, 2001. IEEE Computer Society.